\documentclass[twocolumn,prl,showpacs,preprintnumbers,amsmath,amssymb]{revtex4}
\usepackage{graphicx}
\usepackage{dcolumn}
\usepackage{bm}

\begin{document}

\title{Direct observation of the mass renormalization in SrVO$_3$ by angle
resolved photoemission spectroscopy}

\author{T. Yoshida$^1$, K. Tanaka$^1$, H. Yagi$^1$, A. Ino$^2$, H. Eisaki$^3$, A.
Fujimori$^1$, and Z.-X. Shen$^4$} \affiliation{$^1$Department of
Complexity Science and Engineering and Department of Physics,
University of Tokyo, Kashiwa, Chiba 277-8561, Japan}
\affiliation{$^2$Hiroshima Synchrotron Radiation Center, Hiroshima
University, Kagamiyama 2-313, Higashi-Hiroshima 739-8526, Japan}
\affiliation{$^3$National Institute of Advanced Industrial Science
and Technology, Tsukuba 305-8568, Japan}
\affiliation{$^4$Department of Applied Physics and Stanford
Synchrotron Radiation Laboratory, Stanford University, Stanford,
CA94305}

\date{\today}

\begin{abstract}
We have performed an angle-resolved photoemission study of the
three-dimensional perovskite-type SrVO$_3$. Observed spectral
weight distribution of the coherent part in the momentum space
shows cylindrical Fermi surfaces consisting of the V 3$d$ $t_{2g}$
orbitals as predicted by local-density-approximation (LDA)
band-structure calculation. The observed energy dispersion shows a
moderately enhanced effective mass compared to the LDA results,
corresponding to the effective mass enhancement seen in the
thermodynamic properties. Contributions from the bulk and surface
electronic structures to the observed spectra are discussed based
on model calculations.
\end{abstract}

\maketitle
The effect of electron correlation on the electronic structure of
the transition metal oxides (TMO) has been one of the most
important and fundamental issues in condensed matter physics
\cite{RMPfujimori}. While the high-$T_c$ cuprates and the colossal
magneto resistive (CMR) manganites belong to the charge-transfer
regime of Zaanen-Allen-Sawatzky diagram \cite{ZSA}, light
transition-metal oxides such as perovskite-type Ti and V oxides
are prototypical Mott-Hubbard-type systems. The metal-insulator
transitions (MITs) in Ti and V oxides can be directly compared
with theoretical predictions using the Hubbard model in, e.g.
dynamical mean-field theory (DMFT), and are, therefore, ideal
model systems to study electron correlation phenomena
\cite{Pavarini}. In earlier photoemission studies of
perovskite-type Ti and V oxides, such as Ca$_{1-x}$Sr$_x$VO$_3$
\cite{inouePES, Morikawa} and La$_{1-x}$Sr$_x$TiO$_3$
\cite{Fujimori, Yoshida}, two characteristic structures in the
transition-metal $d$ band have been identified. One is the
coherent part around the Fermi level ($E_F$) corresponding to the
band-like electronic structure, and the other is the incoherent
part 1-2 eV away from $E_F$ corresponding to the remnant of the
lower Hubbard band (LHB), a behavior which is predicted by DMFT
\cite{DMFT}.

In the filling-control Mott-Hubbard system
La$_{1-x}$Sr$_x$TiO$_3$, a critical mass enhancement toward the
MIT has been observed in the electronic specific heats and
magnetic susceptibility \cite{Kumagai}, analogous to the mass
enhancement scenario developed by Brinkman and Rice \cite{GBR}. A
detail analysis of the coherent part of the photoemission spectra
in La$_{1-x}$Sr$_x$TiO$_3$ has revealed to some extent enhancement
of the effective mass when approaching the MIT \cite{Yoshida},
which corresponds to the enhancement of the electronic specific
heat coefficients $\gamma$ toward the MIT critical point
$x\sim0.06$.

On the other hand, Ca$_{1-x}$Sr$_x$VO$_3$ is a bandwidth-control
Mott-Hubbard system. The photoemission results on
Ca$_{1-x}$Sr$_x$VO$_3$ have suggested that the transition from a
correlated metal to the Mott insulating phase is characterized by
spectral weight transfer from the coherent to the incoherent parts
with decreasing $x$\cite{inouePES}. However, their $\gamma$ does
not show such a large change with $x$ and no critical mass
enhancement region is reached in this system \cite{inouePRB}.
Also, the width of the coherent part does not change appreciably
with $x$ while the photoemission spectral intensity at $E_F$
increases as $x$ increases. This behavior was interpreted due to a
strong momentum dependence of the self-energy \cite{inouePES}.
However, a recent ``bulk-sensitive" photoemission study of
SrVO$_3$ and CaVO$_3$ using soft X-rays indicated that the
spectral weight of the coherent part is larger than that reported
in the previous study with low photon energy and that there is no
appreciable spectral weight transfer between SrVO$_3$ and CaVO$_3$
\cite{sekiyama}. A photoemission study using several photon
energies has indicated that the incoherent part includes a certain
amount of contributions from the surface \cite{Maiti}. Theoretical
studies have shown that the surface electronic structure should
have a more strongly renormalized effective mass because of the
reduced co-ordination number and hence the reduced bandwidth
\cite{Liebsch}.

In order to address those fundamental unresolved issues, direct
observation of band dispersion by angle-resolved photoemission
spectroscopy (ARPES) would provide important information about the
mass renormalization. In the present study, we have observed the
band dispersion, the Fermi surface and hence the band
renormalization in the three-dimensional Mott-Hubbard system
SrVO$_3$ by ARPES. Also, we shall address the issue of surface
effects by obtained comparison of the spectra with the calculated
bulk and surface electronic structures.

\begin{figure}
\includegraphics[width=9cm]{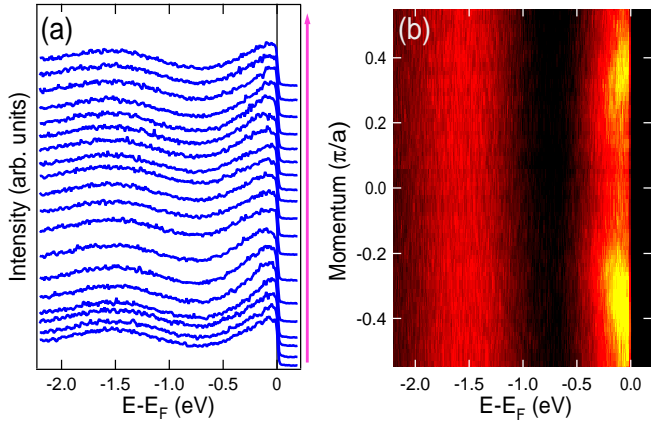}
\caption{\label{incoherent}ARPES spectra for SrVO$_3$ along
momentum cut 4 in the Brillouin zone shown in Fig.
\ref{nkMapping}. (a) EDC's. (b) Intensity plot in the $E$-$k_y$
plane. Dispersive feature within $\sim$0.7 eV of $E_F$ is the
coherent part, while broad feature around -1.5 eV is the
incoherent part.}
\end{figure}

\begin{figure}
\includegraphics[width=8cm]{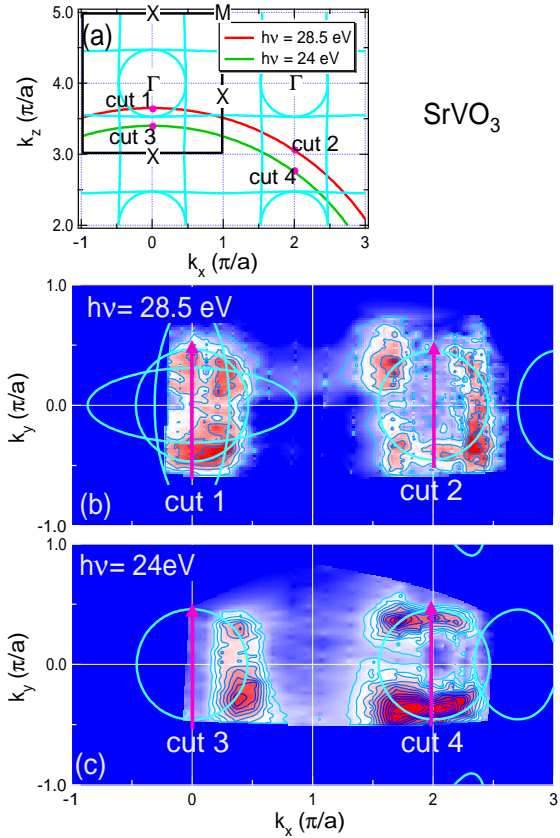}
\caption{\label{nkMapping}Spectral weight mapping at $E_F$. (a)
$k_y$=0 cross-sectional view of the Fermi surfaces (blue curves)
and the momentum loci corresponding to the mapping in panels (b)
and (c). (b) Mapping for $h\nu$=28.5eV. (c) Mapping for $h\nu$= 24
eV. Note that the mapping is projection on the $k_x$-$k_y$ plane.}
\end{figure}

ARPES measurements were performed at Beamline 5-4 of Stanford
Synchrotron Radiation Laboratory (SSRL) with a normal incidence
monochromator and a Scienta SES-200 electron analyzer. The typical
energy and angular resolutions used for the present measurements
were about 30 meV and 0.3 degree, respectively. Single crystals of
SrVO$_3$ were grown using the travelling-solvent floating zone
method. Samples were first aligned by Laue difraction \textit{ex
situ}, and cleaved along the cubic (100) surface and measured at a
temperature of 15 K in a pressure better than $5 \times 10^{-11}$
Torr. We have performed the measurements at photon energies
$h\nu$=24 and 28.5 eV. In this paper, the electron momentum is
expressed in units of $\pi/a$, where $a \sim  3.84$ \AA, is the
lattice constant corresponding to the V-O-V distance. $k_x$ and
$k_y$ are the in-plane momenta and $k_z$ are the out-of-plane
ones.

\begin{figure}
\includegraphics[width=9cm]{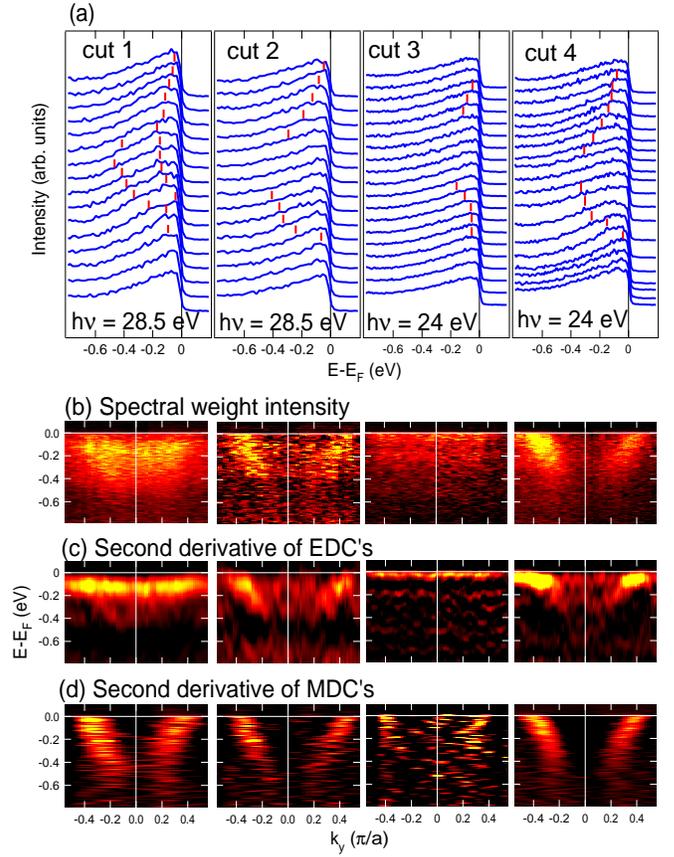}
\caption{\label{EDC_Ek}(Color) Panels (a) show EDC's corresponding
to the cuts in Fig. \ref{nkMapping}. Vertical bars are guides to
the eyes indicating the positions of the dispersive features.
Panels (b) are intensity plots in $E$-$k$ space of the EDC's in
the panels (a). Angle-independent backgrounds have been
subtracted. Second derivatives of the EDC's and MDC's are also
shown in panels (c) and (d), respectively.}
\end{figure}

First, we show an example of energy distibution curves (EDC's) for
SrVO$_3$ in Fig. \ref{incoherent}(a). The coherent part with
$\sim$ 0.7 eV of $E_F$ shows a dispersive feature, consistent with
the view point that the coherent part corresponds to the
band-structure calculation. On the other hand, the incoherent part
centered at $\sim$ -1.5 eV does not show a clear dispersion as
seen in Fig \ref{incoherent}(b). Background from angle-integrated
spectra and the surface effects discussed below may hinder the
observation of subtle dispersive features in the incoherent part.
In the rest of this paper, we shall focus on the dispersive
feature in the coherent part.

Figure \ref{nkMapping} (a) illustrates the $k_y$=0 cross-sectional
view of the Fermi surfaces (blue curves) predicted by
local-density approximation (LDA) band-calculation
\cite{Takegahara} and the loci of electron momenta in the
$k_x$-$k_z$ plane for constant photon energies $h\nu$= 24 eV and
28.5 eV. The inner potential of 10 eV has been assumed. Here, the
LDA band-structure \cite{Takegahara} to a tight-binding (TB) model
consisting of three $t_{2g}$ orbital $d_{xy}$, $d_{yz}$ and
$d_{zx}$ of V. The corresponding three bands do not hybridize with
each other \cite{TBNN} within the present TB formalism which is
the same as that in Ref. \cite{Liebsch}. Figure \ref{nkMapping}
(b) and (c) shows spectral weight mapping in the $k_x$-$k_y$
momentum space at $E_F$ on the momentum loci in panel (a) for the
photon energies $h\nu$=28.5 and 24 eV, respectively. Spectral
weight is integrated over 30 meV within $E_F$. As seen in Fig.
\ref{nkMapping} (b) and (c), spectral weight for both photon
energies are largely confined within the cylindrical Fermi surface
extended in the $k_z$ direction. According to panel (a), cuts 2
and 4 correspond to momenta near the X point
[$\mathbf{k}=(1,0,0)$]. Therefore, the observed spectral weight
comes only from one cylindrical Fermi surface referred to as
$\gamma$ sheet \cite{inoueDHVA} which arises from the $d_{xy}$
orbital. On the other hand, the spectral weight in the first BZ
for $h\nu$ =28.5 eV (cut 1) is more intense than that for 24 eV
(cut 3). This difference comes from the fact that the momentum
locus for cut 1 is within the $\alpha$- and $\beta$ sheets
enclosed by the three cylindrical Fermi surfaces \cite{inoueDHVA},
while that for cut 3 is out of them. Indeed, energy distribution
curves (EDC's) along cut 1 [Fig. \ref{EDC_Ek} (a)] show
complicated features suggestive of two energy dispersions.

Figure \ref{EDC_Ek}(a) shows EDC's along the four cuts in Fig.
\ref{nkMapping}. Panels in Fig. \ref{EDC_Ek}(b) show spectral
weight plots in the energy-momentum ($E$-$k$) space corresponding
to each cut. The angle-independent EDC's have been subtracted as
the background in these $E$-$k_y$ plots \cite{BG}. The second
derivatives of the EDC's and the MDC's are shown in panels (c) and
(d), respectively, so that one can visualize the band dispersions
in spite of the high backgrounds. While cuts 2 and 4 show single
energy dispersions, cut 1, which is around the $\Gamma$ point,
shows two band dispersions as in the EDC's in panel (a). This is
due to the fact that the momenta of cut 1 are within the $\alpha$
and $\beta$ sheets of the Fermi surface.

\begin{figure}
\includegraphics[width=9cm]{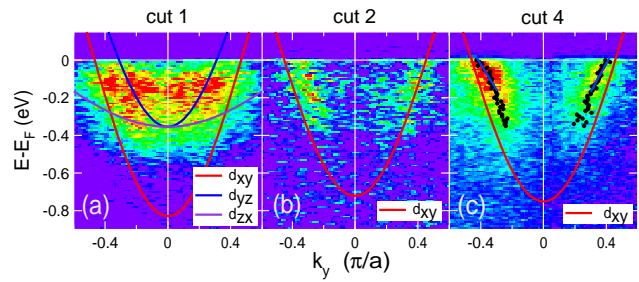}
\caption{\label{Ek_TB}(Color) Comparison of spectral weight
distributions with calculated band dispersions. (a), (b) and (c)
correspond to cuts 1, 2 and 4, respectively. Black dots in (c) are
peak positions of MDC's and represents band dispersion.}
\end{figure}

In Fig. \ref{Ek_TB}, we compare the $E$-$k_y$ plots with the
energy dispersions expected from the LDA calculation. As seen in
the EDC plots of cut 1 and cut 4 in Fig. \ref{EDC_Ek}, there are
parabolic energy dispersions with the bottom at $\sim$-0.5 eV,
which is shallower than that expected from the LDA calculation
$\sim$ -0.9 eV. The two dispersions observed for cut 1 in Fig.
\ref{EDC_Ek} may correspond to the dispersions of the $d_{xy}$ and
$d_{yz}$ bands, as shown in Fig. \ref{Ek_TB} (a). The dispersion
of $d_{zx}$ is not clearly observed in the present plot, probably
due to the effect of transition matrix-elements. For cuts 2 and 4
[Fig. \ref{Ek_TB} (b) and (c)], since they do not intersect the
$d_{yz}$ and $d_{zx}$ bands, we observed only the $d_{xy}$ band.

In panel (c), we have derived the band dispersion for cut 4 from
MDC peak positions indicated by black dots. By comparing the Fermi
velocity of the LDA band structure and that of the present
experiment, one can see the overall band narrowing in the measured
band dispersion. Near $E_F$, we obtain the mass enhancement factor
$m^*/m_b \sim 1.8\pm 0.2$, which is close to the value $m^*/m_b=
1.98$ obtained from the specific heat coefficient $\gamma$
\cite{inouePRB}. Since the LDA calculation predicted that the
$d_{xy}$ band is highly two-dimensional and isotropic within the
$k_x$-$k_y$ plane, a similar mass enhancement is expected on the
entire Fermi surface.

It has been pointed out that the photoemission spectra of this
system taken at low photon energies have a significant amount of
contributions from surface states, particularly in the incoherent
part centered at $\sim$ -1.5 eV \cite{sekiyama, Maiti}. In the
photon energy range used in the present work, indeed, the bulk
spectra has a smaller spectral weight compared to that of surface,
by a factor of $\sim$2 \cite{Maiti}. However, surface spectra tend
to have spectral weight in the incoherent part, reducing their
spectral weight near $E_F$. Therefore, surface contributions are
thought to be relatively small in the coherent part but may not be
negligible. In order to distinguish bulk from surface
contributions in the present spectra, we shall discuss the
character of surface states.

Here, we consider several possible origins of surface states. One
is a surface reconstruction, which may create two dimensional
electronic states on the first layer different from bulk states.
Such reconstruction-derived surface bands were observed in the
ARPES spectra of Sr$_2$RuO$_4$. In that case, the surface band was
symmetric with respect to the ($\pi$,0)-(0,$\pi$) line due to band
folding into the new Brillouin zone of the
$\sqrt{2}\times\sqrt{2}$ superstructure. Since such a folded band
has not been observed in the present results, the effect of
surface reconstruction would not exist or would be negligible in
the present ARPES spectra.

\begin{figure}
\includegraphics[width=9cm]{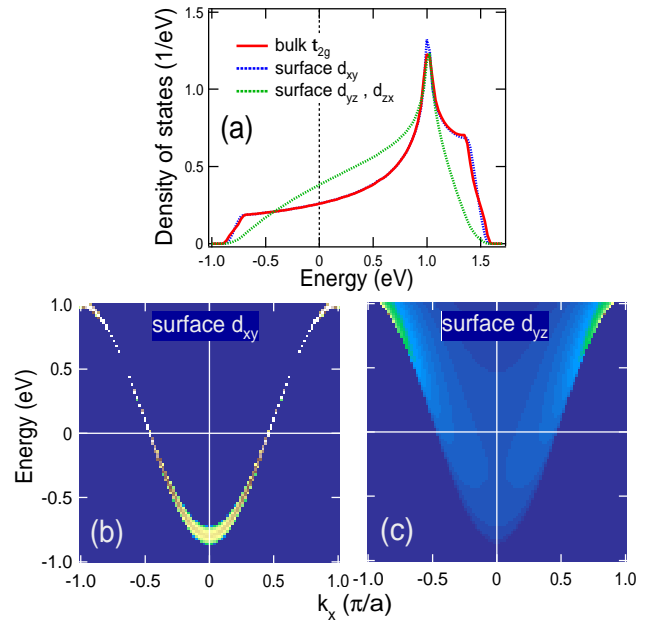}
\caption{\label{surface_states} Surface-projected electronic
states calculated using the semi-infinite tight-binding model. (a)
Density of states. (b) Spectral weight distribution of $d_{xy}$
states. (c) Spectral weight distribution of $d_{yz}$ states.}
\end{figure}

Another possible origin of surface states is the discontinuity of
the potential at the surfaces. Figure \ref{surface_states} shows
surface-projected density of states and band dispersion states
calculated using a tight-binding Green's function formalism
\cite{Kalstein} for a semi-infinite system in the same manner as
Ref. \cite{Liebsch}. As shown in panels (a) and (b), the surface
$d_{xy}$ band does not show an appreciable change from the bulk
band because it has the two-dimensional electronic structure
within the $x$-$y$ plane. In contrast, spectral weight of the
surface $d_{yz, zx}$ states is redistributed between the bulk
$d_{yz, zx}$ band and the Fermi level [Fig.
\ref{surface_states}(c)], because $k_z$ is no longer a good
quantum number. This indicates that we cannot observe clear
dispersions of the surface $d_{yz, zx}$ bands in ARPES.
Accordingly, the observed dispersion of the $d_{xy}$ band should
be similar to that of the bulk $d_{xy}$ band and would therefore
represent the bulk electronic states rather than the surface
states.

In conclusion, we have directly observed the energy dispersions
and the Fermi surfaces of SrVO$_3$ by ARPES. The observed
effective mass renormalization in the $d_{xy}$ band near $E_F$ is
consistent with the moderate mass enhancement in $\gamma$. From
the model calculations of the surface electronic structure, we
conclude that the observed dispersion of the $d_{xy}$ band
represents the bulk electronic structure. The present study has
demonstrated that ARPES measurements are useful even for
three-dimensional perovskite systems and provide direct
information about the band renormalization in Mott-Hubbard system
near MIT's.

We are grateful to M. Rozenberg and I. H. Inoue for enlightening
discussions and D. H. Lu for technical support. This work was
supported by a Grant-in-Aid for Scientific Research ``Invention of
Anomalous Quantum Materials" from the Ministry of Education,
Science, Culture, Sports and Technology, Japan. SSRL is operated
by the Department of Energy's Office of Basic Energy Science,
Division of Chemical Sciences and Material Sciences.


\begin{thebibliography}{999}
\bibitem{TBNN} Distant-neighbor hopping couses hybridization
between the three $t_{2g}$ bands. However, this has small effect
and may be a negligible in the present analysis.
\bibitem{BG} For the present analysis, we have used EDC's with no dispersive feature
outside of the cylindrical Fermi surface as a background. The high
background seen in the EDC's may come from the angle-integrated
spectral weight due to disorder or inperfection of the irregularly
cleaved or fractured part of the surfaces. Nonetheless, we could
observe clear structures with energy dispersions on the
background. We consider that the dispersive features come from the
well cleaved part of the sample surface.
\end{thebibliography}
\end{document}